\documentclass[aps,twocolumn,floats,prd,nofootinbib,superscriptaddress,10pt]{revtex4-1}

\usepackage[dvips]{graphicx} %
\usepackage{graphicx,amsmath,amsfonts,amssymb,slashed}
\usepackage{bbold,wasysym}
\usepackage{graphicx}

\usepackage[usenames,dvipsnames]{xcolor} 

\usepackage{soul}

\definecolor{RedWine}{rgb}{0.743,0,0}
\definecolor{RoyalBlue}{rgb}{0.25,.41,.88}

\setstcolor{Blue}

    \newcommand{\be}{\begin{equation}}
\newcommand{\ee}{\end{equation}}
\newcommand{\ba}{\begin{align}}
\newcommand{\ea}{\end{align}}

\newcommand{\VEV}[1]{\left<{1}\right>}

\begin{document}

\title{Large-Distance Lens Uncertainties and Time-Delay Measurements of $H_0$}

\author{Julian B.~Mu\~noz} 
\affiliation{Department of Physics and Astronomy, Johns Hopkins
	University, 3400 N.\ Charles St., Baltimore, MD 21218, USA}
\affiliation{Department of Physics, Harvard University, Cambridge, MA 02138}
\author{Marc Kamionkowski}
\affiliation{Department of Physics and Astronomy, Johns Hopkins
     University, 3400 N.\ Charles St., Baltimore, MD 21218, USA}

\begin{abstract}
Given the tension between the values of the Hubble parameter
$H_0$ inferred from the cosmic microwave background (CMB) and
from supernovae, attention is turning to time delays of strongly
lensed quasars.  Current time-delay measurements indicate a
value of $H_0$ closer to that from supernovae, with errors on the
order of a few percent, and future measurements aim to bring the
errors down to the subpercent level.  Here we consider the
uncertainties in the mass distribution in the outskirts of the
lens.  We show that these can lead to errors in the inferred
$H_0$ on the order of a percent and, once accounted for,
would correct $H_0$ upward (thus increasing slightly the tension with
the CMB).  Weak gravitational lensing and simulations may help
to reduce these uncertainties.
\end{abstract}

\maketitle

The Hubble tension may now well provide the greatest challenge
to the canonical cosmological model
\cite{Freedman:2017yms,Addison:2017fdm}.  The value of the
Hubble parameter $H_0$ obtained from the cosmic
microwave background (CMB), where $H_0$ affects the very
precisely determined angular scale of the acoustic peaks in the
CMB power spectrum
\cite{Jungman:1995bz,Hu:1996vq,Aubourg:2014yra}, is
$H_0^{\rm CMB}=67.3 \pm 1.0$ km~s$^{-1}$~Mpc$^{-1}$
\cite{Ade:2015xua}, which tightens to $H_0^{\rm CMB + gal}=67.6
\pm 0.6$ km~s$^{-1}$~Mpc$^{-1}$ when supplemented by galaxy-survey data
\cite{Seo:2003pu,Beutler:2011hx,Ross:2014qpa,Anderson:2013zyy}.

The Hubble parameter can also be obtained by
comparing the brightnesses and redshifts of standard candles
\cite{Hubble:1929ig}.  Recent supernova observations have
determined the value of the Hubble parameter to be $H_0^{\rm
SNe}=73.2 \pm 1.7$ km s$^{-1}$ Mpc$^{-1}$~\cite{Riess:2016jrr}, at roughly $3$-$\sigma$
tension with the CMB-inferred value.  
Cosmological explanations
of the discrepancy are not easily come by; they typically involve
some modifications to the cosmic expansion history that then
introduces some other tension with the detailed structure of the
CMB power spectrum
\cite{Riess:2016jrr,Bernal:2016gxb,Karwal:2016vyq}. Another
possibility is that the discrepancy may arise from measurement
biases in one or both observables
\cite{Efstathiou:2013via,Addison:2015wyg,Dhawan:2017ywl}, and so it is of
paramount importance to obtain a third independent probe of
$H_0$. 

Attention is thus turning now to the value of $H_0$ inferred
from time delays of strongly lensed quasars
\cite{Refsdal:1964nw,Kochanek:2002rk,Oguri:2006qp,Treu:2016ljm}.
There has been tremendous recent progress in this endeavor, with
the H0LiCOW program recently reporting $H_0^{\rm lens}=71.9^{+2.4}_{-3.0}$
km~s$^{-1}$~Mpc$^{-1}$ \cite{Bonvin:2016crt} from three lensing
systems.  Additional lenses are expected to reduce the error
bars on $H_0$ even further \cite{Suyu:2016qxx}.

In order to reach not only percent-level precision, but also
percent-level accuracy, the mass distribution of the lens must
be carefully modeled.  For example, uncertainties in the radial
mass profile assumed for the lens have been shown to induce
errors of several percent in $H_0$
\cite{Schneider:2013sxa,Xu:2015dra,Tagore:2017pir}; 
whereas microlensing of the quasar source can cause comparable uncertainties \cite{Kochanek:2017}.
Here, we
focus specifically on the mass distribution of the lens at large
distances from the lens center of mass.  We show that
uncertainties in this large-distance mass distribution may lead
to uncertainties in $H_0$ of a few percent.  We also argue that
current modeling may be biasing the value of $H_0$ down
(implying greater tension with the CMB).

The mass distribution of lens galaxies at large radii remains, to a
great extent, unknown.  Galactic mass profiles must be truncated
at no more than the $\sim$Mpc typical intergalactic spacing, and
weak-lensing studies \cite{Brimioulle:2013kva} suggest the
mass distributions of galaxies that resemble typical strong lenses ought to be truncated at distances $\lesssim
500$~kpc.  The time delay depends on the total mass projected
along the line of sight \cite{Kochanek:2003pi}, and so there may
be artificial contributions to the expected time delay if the
lens mass is not truncated.  Although this effect was too small to
be of concern in prior work, it introduces, if neglected, a
$\mathcal O(1\%)$ bias in the inferred Hubble parameter and, if
considered, still implies a residual uncertainty of comparable
magnitude.

We also consider a subtle issue about how cosmological lenses
are embedded in an FRW universe.  The usual discussions of
lensing surmise that the mass associated with a lens is {\it
added} to an otherwise homogeneous FRW universe, giving rise to
a potential perturbation that falls off as $1/r$ with distance
$r$ from the lens.  In our Universe, however, the mass
associated with any given lens arises from a local overdensity
which is compensated elsewhere with an underdensity.  As a
result, the potential perturbation associated with any
particular lens should fall off far more rapidly than $1/r$ at
large distances.  We show that a correct accounting of this
effect biases the inferred Hubble parameter downwards, but only
by $(\delta H_0/H_0)\sim10^{-4}$.

We begin by reviewing
the lensing formalism.  Given a mass density $\rho(\mathbf r)$
of the lens, the mass distribution projected onto the
lens plane at angular position $\pmb \theta$ is obtained by
integrating over the line-of-sight distance $z$,
\be
\Sigma (\pmb\theta) = \int dz\, \rho(D_L \pmb \theta,z),
\label{eq:SigmaLens}
\ee
where $D_L$ is the angular-diameter distance to the lens.
We can divide $\Sigma$ by the critical density $\Sigma_{\rm
crit} \equiv c^2 \, D_S/(4\pi G \,D_L D_{LS})$ to separate strong from weak lensing, where $D_S$ and $D_{LS}$
are, respectively, the angular-diameter distances to the source
and between the lens and the source, to obtain the
convergence~\cite{Narayan:1996ba} $\kappa(\pmb \theta) = \Sigma
(\pmb\theta)/\Sigma_{\rm crit}$.
The lensing potential is the projection of the gravitational
potential $\phi$, given by \cite{SchneiderBook},
\be
\psi (\pmb\theta) = \dfrac{2 D_{LS} }{c^2\, D_L D_S} \int dz\, \phi(D_L \pmb \theta,z),
\ee
which is related to the convergence through $\nabla^2_\theta \psi = 2 \kappa$.
This potential yields a deflection angle $\pmb \alpha = \pmb
\nabla_\theta \psi$, which defines where images are formed
through the lens equation,
\be
\pmb\beta = \pmb \theta - \pmb\alpha (\pmb \theta),
\label{eq:lenseq}
\ee
where $\pmb\beta$ is the impact parameter.
The $\pmb\beta$ in Eq.~\eqref{eq:lenseq} is unknown {\it a
priori}, and is obtained by fitting to the observed image
positions $\pmb \theta$, and the $\pmb\alpha(\pmb \theta)$ is
predicted by the lens model (i.e., mass distribution). Signals
from the source will arrive as different images, at
positions $\pmb\theta_i$ and $\pmb\theta_j$, with a time delay
given by \cite{Schneider:1985,Blandford:1986zz}
\be
\Delta t_{ij} =  \dfrac{D_{\Delta t}}{c} \left ( \dfrac{\pmb\alpha^2(\pmb\theta_i)-\pmb\alpha^2(\pmb\theta_j)}{2} - \left[\psi(\pmb\theta_i)-\psi(\pmb\theta_j)\right]\right),
\label{eq:Deltat}
\ee
where we have defined the time-delay distance,
\be
D_{\Delta t} \equiv (1+z_L) \dfrac{D_L D_S}{D_{LS}} \propto H_0^{-1},
\ee
and $z_L$ is the redshift of the lens.  Given that $D_{\Delta
t}$ is a ratio of distances, it is inversely proportional to the
Hubble parameter $H_0$ and only weakly dependent on other
cosmological parameters \cite{Suyu:2016qxx}.

The usual procedure is to consider a parametrized family of
convergences $\kappa(\pmb \theta; \pmb \xi)$ with parameters
$\pmb \xi$.  These parameters are obtained by fitting to
the observed image positions $\{\pmb\theta_i\}$. The Hubble
parameter is then inferred by comparing the time delay expected
from Eq.~\eqref{eq:Deltat}  with that observed.

One issue that arises, though, is the mass-sheet degeneracy,
in which the effect of a constant additional surface-mass
density on the observed image positions can be compensated by a
change in the impact parameter.  If the real convergence of the
lens is $\kappa^{\rm  real}$, but it is modeled as
\be
\kappa^{\rm model} = (1-\lambda) \kappa^{\rm real} + \lambda,
\label{eq:MSkappa}
\ee
the observed image positions will be the same, granted that the
impact parameter is changed as $\pmb \beta^{\rm model} =
(1-\lambda) \pmb \beta^{\rm real}$.
However, the expected time delay is changed to $\Delta
t_{ij}^{\rm model} = (1-\lambda) \Delta t_{ij}^{\rm real}$,
thus yielding a different value,
\be
H_0^{\rm model} = (1-\lambda) H_0^{\rm real},
\label{eq:MSH0}
\ee
for the Hubble parameter.  The mass-sheet degeneracy is not just
a theoretical curiosity. It is expected that the large-scale
structure along the line of sight causes light rays to focus and
defocus, introducing an external convergence $\kappa_{\rm ext}$
\cite{Seljak:1994wa,Rusu:2017}.

There are two avenues to breaking this degeneracy.  The first is to
use dynamical measurements of stellar velocities in the lens, as
the transformation in Eq.~\eqref{eq:MSkappa} also implies a
change $(\sigma_{\rm vel}^{\rm model})^2 = (1-\lambda)
(\sigma_{\rm vel}^{\rm real})^2$ to stellar velocities \cite{Jee:2014uxa}.
In practice, however, uncertainties in the lens profile can
induce errors when extrapolating the mass measurement at small
radii to the larger Einstein radius
\cite{Suyu:2012rh}. Moreover, the possibility of anisotropy in
the lens hampers translation from kinematic data to the
lens mass \cite{Schneider:2013sxa}.

The second method is to simulate fields of view in cosmological
$N$-body simulations to obtain a probability distribution function
(PDF) for $\kappa_{\rm ext}$ \cite{Hilbert:2008kb,Suyu:2012aa,Birrer:2016xku}.
In an FRW universe this PDF has mean $\left<\kappa_{\rm
ext}\right>=0$, but the finite width $\left<\kappa_{\rm
ext}^2\right>\neq0$ is one of the limiting factors in
current time-delay $H_0$ measurements \cite{Suyu:2012aa}.  There
has been great development in the study of this PDF; for
instance, we have learned that multiply imaged quasars, as
biased tracers of the underlying matter distribution, live
preferentially in overdense regions, which causes a
percent-level bias on $\left<\kappa_{\rm ext}\right>$
and thus on the inferred $H_0$ \cite{Collett:2016muz}.  An
example of this bias is found in the lens system RXJ1131-1231,
which resides in a line of sight with $\sim 40\%$ more galaxies
than average \cite{Suyu:2012aa}, which causes the expectation
value of the external convergence to be $\left<\kappa_{\rm
ext}\right>\approx 0.1$.  In an effort to find the PDF of
$\kappa_{\rm ext}$ for each individual system, instead of the
average PDF of an FRW universe, both the average number counts of
galaxies in the field \cite{Suyu:2009by}, as well as the
external shear $\gamma_{\rm ext}$ \cite{Suyu:2012aa}, have been
used as ancillary data.

The aforementioned $N$-body studies quantify the contributions of
independent structures, along the line of sight, that are at
large (cosmological) physical distances from the lens.  What we
will consider now, though, is the mass distribution {\it in the
lensing system}, but at physical distances (e.g., $\sim100$s
kpc) large compared with the Einstein radius and impact
parameter (e.g., $\sim10$ kpc).  We will first show that this
can be approximated as a mass-sheet transformation.

We focus on the spherically symmetric power-law models,
with mass density given by 
\be
\rho_{\rm model} (r) = \rho_0 \left( r/r_0 \right)^{-\gamma'},
\label{eq:rhoPL}
\ee
that are usually used in lens modeling.  This density profile
gives rise to a projected surface-mass density at a distance
$b(= D_L\theta)$ from the center of the lens of
\be
\Sigma_{\rm model} (b) = \sqrt{\pi} \rho_0 r_0^{\gamma'}
b^{1-\gamma'} \Gamma\left(\frac{\gamma'-1}{2}\right)
\left[\Gamma\left(\frac{\gamma'}{2}\right) \right]^{-1},
\ee
where $\Gamma$ is the gamma function.
We now compute the critical density through $\pi R_E^2
\Sigma_{\rm crit} = M_{\rm los}$, where $M_{\rm los}$ is the
line-of-sight mass contained within a cylinder of radius $R_E$,
to find \cite{Suyu:2009by}
\be
\Sigma_{\rm crit} = -\rho_0 r_0^{\gamma'} R_E^{1-\gamma'}
\sqrt{\pi} \,
\Gamma\left(\frac{\gamma'-3}{2}\right)
\left[\Gamma\left(\frac{\gamma'}{2}\right)\right]^{-1}.
\label{eq:SigmaCrit}
\ee
We then obtain the convergence \cite{Barkana:1998qu,Jee:2014uxa}
$\kappa_{\rm model} (\theta) = (3-\gamma') \left(
\theta_E/\theta \right)^{\gamma'-1}/2$.
The parameters of the model are the power-law index $\gamma'$,
and the Einstein angle $\theta_E\equiv R_E/D_L$.  Using this
model, augmented with an ellipticity parameter to account for
the noncircularity of the lens, the authors of Ref.~\cite{Suyu:2012aa}
inferred a Hubble constant $H_0= 78.7^{+4.3}_{-4.5}$ km s$^{-1}$ Mpc$^{-1}$ from the
RXJ1131-1231 system.

The issue we first address is the uncertainty associated with
the assumption of a power-law mass distribution that extends to
infinite radius.  It is clear that the mass distribution cannot
extend to infinity (and that the total mass cannot be infinite,
as the power-law mass profile implies).  Still, the contribution
to the convergence, and thus the observables, is small enough to
be neglected in prior work.  As we move to subpercent
precision/accuracy, though, the effects of the truncation radius
become significant.  To see this, we truncate the mass density
from Eq.~\eqref{eq:rhoPL} at a finite radius by adding the
negative-mass distribution $
\rho_t = - \rho_0 \left(r/r_0\right)^{-\gamma'} \Theta(r-r_t)$,
to the model,
where $\Theta(r)$ is the Heaviside step function, and $r_t$ is
the truncation radius.  This distribution gives rise to a
projected surface mass density $ \Sigma_{t} = - 2 \rho_0
r_0^{\gamma'} r_t^{1-\gamma'}/(1-\gamma')$,
neglecting terms of $\mathcal O[(b/r_t)^3]$ or larger.
Again dividing by the critical density\footnote{Which is
accurate to first nonvanishing order in $R_E/r_t$, since
otherwise $\Sigma_{\rm crit}$ from Eq.~\eqref{eq:SigmaCrit}
would depend on $r_t$.} from
Eq.~\eqref{eq:SigmaCrit},
we find the convergence due to this negative-mass distribution to be
\be
\kappa_t = \dfrac{2 \, \Gamma\left(\frac{\gamma'}{2}\right) (\gamma'-1)}{\sqrt{\pi} \, \Gamma\left(\frac{\gamma'-3}{2}\right)} 
\left(\dfrac{R_E}{r_t}\right)^{\gamma'-1} < 0.
\label{eq:kappatgamma}
\ee
This large-radius negative-mass
distribution thus modifies the convergence to
\be
\kappa_{\rm real} (\theta) = \dfrac{3-\gamma'}{2} \left( \dfrac{\theta_E'}{\theta} \right)^{\gamma'-1} + \kappa_t,
\label{eq:kappareal}
\ee
independently of $\theta$, and is then equivalent to 
a mass-sheet transformation with $\lambda_t = -\kappa_t$. 
We thus use Eq.~\eqref{eq:MSH0} to relate the real $H_0$ to the
one inferred by the nontruncated model,
\be
H_0^{\rm real} = \dfrac{H_0^{\rm model}}{(1+\kappa_t)} \approx (1-\kappa_t) H_0^{\rm model} > H_0^{\rm model}.
\label{eq:truncH0}
\ee
Thus, time-delay measurements of $H_0$ are biased low if the
finite extent of the lensing mass distribution is not taken into
account.

We now consider the range of reasonable values for the
truncation radius $r_t$.  As the analysis above indicates, the
image positions do not depend significantly on the mass
distribution at large radii, a consequence of the fact that the
light rays for observed images have trajectories with impact
parameters comparable to the Einstein radius, which is much
smaller than $r_t$ \cite{Tagore:2017pir}.  Nonetheless, galaxies
produce weak lensing at very wide angular separations, which can
be detected by the shear created on background galaxies
\cite{Bartelmann:1999yn}.  In Ref.~\cite{Brainerd:1995da} a
study of the truncation radius of galaxies was performed, where
the lens mass density was modeled as a dual pseudo isothermal elliptical (dPIE) mass distribution~\cite{Eliasdottir:2007md}
\be
\rho(r) = \dfrac{\sigma_{\rm vel}^2 s^2}{4\pi G\, r^2 (r^2+s^2)},
\label{eq:rhos}
\ee
which is isothermal for $r\ll s$, and decays as $r^{-4}$ for
larger radii, effectively showing a cutoff at $r \sim s$.  To
first nonvanishing order in $r/s$ the mass distribution in
Eq.~\eqref{eq:rhos} yields a convergence
\be
\kappa(r) = \dfrac{\theta_E}{2\theta} \left( 1 + \dfrac{R_E}{2 s} \right) - \dfrac{R_E}{2s},
\ee
which can be identified with the isothermal case ($\gamma'=2$) of our
Eq.~\eqref{eq:kappareal}, if
$\kappa_t = - {R_E}/({2 s})$.
By comparing to Eq.~\eqref{eq:kappatgamma}, $\kappa_t = -
R_E/(\pi r_t)$, we find $s = \pi r_t/2$.
This also shows that for the purposes of strong lensing, where
$r \sim R_E \ll r_t$, our sharp cutoff is a good approximation
to the smooth truncation scheme in Eq.~\eqref{eq:rhos}, while
remaining valid for $\gamma'\neq2$, thus fitting most lens models, which are not isothermal.  The size of $s$ has been
estimated in Ref.~\cite{Brainerd:1995da} to be $s\gtrsim 100
h^{-1}$ kpc, whereas a more recent study in
Ref.~\cite{Hoekstra:2003pn} found $s = 185^{+30}_{ - 28}\, h^{-1}$
kpc on average over an ensemble population of all galaxies. 
Furthermore, in Ref.~\cite{Brimioulle:2013kva} it was found that
red galaxies, which tend to be early type and thus more likely
to be strong lenses, have on average larger truncation radii, $s
\approx 300\, h^{-1}$ kpc. 

We thus find that time-delay Hubble-parameter measurements are
biased low by
\be
   \frac{\delta H_0}{H_0} \approx - 0.01 \left(
   \frac{R_E}{10\,{\mathrm{kpc}}} \right) \left(
   \frac{r_t}{300\,{\mathrm{kpc}}} \right)^{-1},
\ee
where for simplicity we have set $\gamma'=2$.
Although the precise bias will differ from lens to lens, the
bias will survive even if Hubble parameters inferred from multiple
systems are averaged, as it has the same sign for all lenses,
and thus averages to some nonzero value $\bar\kappa_t$.
The uncertainties in the values of $r_t$ for each lens introduce
moreover an accompanying error in the inferred value of $H_0$.
If the $\kappa_t$ for different lenses are distributed
about the mean with a variance $\sigma^2_{\kappa_t}$, then there
will still be an uncertainty in $H_0$ of $\sigma_{H_0}/H_0
\approx  \sigma_{\kappa_t}/\sqrt{N}$, from $N$ time-delay systems.
Moreover, the average $\bar\kappa_t$ between the lenses can only be inferred with an error $\sigma_{\bar\kappa_t}\approx \sigma_{\kappa_t}/\sqrt{N}$. 
Therefore, subtracting our estimate of the average from the data yields a residual bias $\delta H_0/H_0 \sim \sigma_{\bar\kappa_t}$.
Detailed studies of the lens-galaxy population are thus
imperative to overcome these uncertainties.

We have chosen a simple power-law model to illustrate the
effects of truncation of the mass distribution, although a similar
uncertainty should be present in other models.  
For instance, in Refs.~\cite{Baltz:2007vq,Frittelli:2011uh} truncated NFW models were presented.
Furthermore, in Refs.~\cite{Suyu:2013kha,Wong:2016dpo} the lens
systems RXJ1131-1231 and HE 0435−1223
were fit with both a power-law distribution
and a composite model, which includes dark matter and baryons.
The composite model presents an effective cutoff with respect to
the power-law, due to the faster decrease of the dark-matter
density at large radii~\cite{Schneider:2013sxa}.  This is to be compared with our modeled
cutoff in Eq.~\eqref{eq:kappareal}, from which we would expect a
higher inferred value of $H_0$ for the composite model.
Nevertheless, this effect---which we estimate to bias $H_0$ by one percent---is smaller than current measurement and modeling uncertainties.

There is another issue, of a more conceptual nature, which we now
consider.  In the usual discussions of lensing, a lensing mass
distribution (e.g., from a galaxy or cluster) is {\it added} to an
otherwise FRW universe, thus giving rise to potential
perturbations that fall off as $1/r$ with the distance $r$ from
the lens.  In our Universe, however, galaxies and clusters are
formed from local overdensities, in an otherwise FRW universe,
that are then compensated by underdensities elsewhere.  Thus, if
we go to distances large compared with the typical intergalactic
separation, there will be no residual $1/r$ potential
perturbation (somewhat analogously to Debye
screening in a plasma).  
What we are considering here thus {\it
differs} from prior work  \cite{Piattella:2015xga,Chen:2010gi}
in which the lens was embedded in a spacetime that asymptotes to
FRW at large distance (the residual $1/r$ potential perturbation
still arises there).
Our analysis also differs from that of Ref.~\cite{McCully:2016yfe} in that we compensate the mass of the strong lens, instead of the weak perturbers along the line of sight.

To estimate the impact of this issue, we consider a lens
of mass $M$ that is surrounded by a spherical negative-mass
shell (NMS) of same total mass at some large radius
$R_f$---i.e., we take the lens to be a spherical mass
distribution of zero total mass.  We take $R_f$ to be the radius
in a homogeneous universe of matter density $\rho_m$, at which
an object of mass $M_L$ dominates the gravitational potential;
i.e., $R_f \sim [3\,M_L/(4\pi\rho_m)]^{1/3}$, which for a
matter density of $\rho_m (z) \approx 5 \times 10^{-8} (1+z)^3
\,M_\odot$/pc$^3$ is 
\be
     R_f \sim \,{\rm Mpc}\, \left(
     \dfrac{M_L}{10^{11}M_\odot}\right)^{1/3} \dfrac 1 {1+z_L}.
\ee
The NMS has a mass distribution,
\be
\rho_{\rm NMS} = - \dfrac{M_g}{4\pi R_f^2} \delta_D(r-R_f),
\label{eq:NMSrho}
\ee
which gives rise to a convergence,
\be
\kappa_{\rm NMS} (b) = -\dfrac{R_E^2}{2 R_f \sqrt{R_f^2 - b^2}}.
\ee
For $R_f \sim$~Mpc, and for an Einstein radius $R_E\sim 10$ kpc, we
find $ \kappa_{\rm NMS} \approx - (R_E^2/R_f^2)/2 \sim  - 10^{-4}$,
which is, again, independent of angle to first nonvanishing
order.  The convergence thus resembles a
negative-mass-sheet, since we are only observing it at distances
$b \ll R_f$, where the curvature of the NMS is negligible.  The
magnitude of the bias and uncertainty introduced in $H_0$
measurements is only $\mathcal O(10^{-4})$ and thus not
significant for current or forthcoming measurements of $H_0$.

We now return to the bias and error in $H_0$ introduced
by the uncertainty in the large-radius mass distribution, and now
consider what is known about the truncation radius and
what more might be learned about it in the future.
Weak-lensing measurements are already beginning to provide some
constraints on the average value of $r_t$, but $r_t$ varies
amongst different types of galaxies~\cite{Brimioulle:2013kva}.
It will thus be important to extend such measurements further
restricting the population of lens galaxies to those that more
closely resemble strong-lensing systems.  The challenge here
will be statistics with the reduced number of systems and then
beyond that, separating the effects, in galaxy-galaxy lensing,
of the lens potential, from those of large-scale clustering.

Even if a lens-like population of galaxies can be well characterized, one might want to measure the truncation radius for an individual lens.
This will be difficult with traditional weak-lensing measurements,
given the relatively small masses of the lens galaxies and the
finite number of background sources to be lensed.  Still, in the
longer term, radio arrays may provide measurements of the 21-cm
line during the dark ages to arcsecond resolution. This would
allow studies of weak lensing around individual objects,
characterizing their environment to great accuracy~\cite{Kovetz:2012jq}.


Although simulations might not
shed light onto the large-distance mass distribution of every individual lens, there is more that can be done to determine the PDF of the effective convergence associated with a family of lenses. The procedure should be analogous to that used to infer the PDF of the external convergence due to line-of-sight objects, (e.~g., ray-tracing through simulations,)
albeit applied to the outskirts of lens-like galaxies with the necessary resolution.

As yet, the effects we have considered here have been
subdominant compared with other uncertainties associated with
modeling the lens mass distribution.  As we move forward,
though, to subpercent precision, there will need to be more
focus on the mass distribution in the outskirts of the lens,
work that can be pushed forward with weak lensing and simulations, 
to enable a precise and unbiased subpercent-level measurement
of $H_0$.

\begin{acknowledgments}
The authors thank Y.\ Ali-Ha\"imoud, I.\ Bah, R.\ 
Caldwell, L.\ Dai, C.\ Dvorkin, E.\ Kovetz, T.\ Smith, S.\ Suyu, T.\ Venumadhav, and D.\ Weinberg for useful discussions.  This work was supported at JHU
by NSF Grant No.\ 0244990, NASA NNX15AB18G, and the Simons
Foundation.
\end{acknowledgments}

\end{document}